\documentclass[11pt,aps,showpacs,showkeys,preprintnumbers,amsmath,amssymb,nofootinbib]{revtex4}

\usepackage{subfigure}
\usepackage{graphicx}
\usepackage{color}
\usepackage{amsmath}
\usepackage{amsfonts}
\usepackage{amssymb}
\usepackage{adjustbox}
\def \be  {\begin{equation}}
\def \ee  {\end{equation}}
\def \ee  {\end{equation}}
\def \bea {\begin{eqnarray}}
\def \eea {\end{eqnarray}}

\providecommand\textprime{}

\begin{document}

\vspace*{1mm}

\title{Particle Ratios within a statistically corrected hadron resonance gas model and EPOS event-generator at AGS, SPS, RHIC and LHC Energies}

\author{Mahmoud Hanafy}
\email{mahmoud.nasar@fsc.bu.edu.eg}
\affiliation{Physics Department, Faculty of Science, Benha University, Benha, Egypt}

\author{Abdel Nasser Tawfik}
\email{a.tawfik@fue.edu.eg}
\affiliation{Future University in Egypt (FUE), 11835 New Cairo, Egypt}

\author{Muhammad Maher}
\email{m.maher@science.helwan.edu.eg} 
\affiliation{Helwan University, Faculty of Science, Physics Department, Ain Helwan, Egypt}

\date{\today}

\begin{abstract}

We further investigate the applicability of our previously suggested quantum-mechanically correlated statistical hadron gas model (HRG) to the ideal hadron resonance gas model (IHRG), which is inspired by a Beth-Uhlenbeck corrected form of the equation of state (EoS). We compute the ratios of several particle yields, both equal-mass pairs ($\bar{p} / p$, $K^-/ K^+$, $\pi^-/ \pi^+$, $\bar{\Lambda}/ \Lambda$, $\bar{\Sigma}/ \Sigma$, $\bar{\Omega}/ \Omega$) and unequal-mass pairs ($p/ \pi^+$, $k^+/ \pi^+$, $k^-/ \pi^-$, $\Lambda/ \pi^-$, $\bar{p}/ \pi^-$, $\Omega/\pi^-$) and investigate how these ratios change with the center-of-mass energy. Next, we present a comparative analysis of the outputs of our suggested HRG model and the IHRG model, Cosmic Ray Monte Carlo (CRMC) EPOS $1.99$ simulations, and experimental data from ALICE, SPS, AGS, and RHIC. When compared to the other models taken into consideration, our HRG model typically shows very close agreement with the experimental results. The proton anomaly reported at top RHIC and LHC energies may be addressed by our new HRG model, which notably exhibits a strong alignment with experimental data for $\bar{p}/ \pi^-$ and $p/ \pi^+$ ratios. However, both our HRG model and the IHRG model seem to significantly underestimate some experimental data for ratios involving hadron couples with uneven mass and (multi)strange content, such as $\Lambda/ \pi^-$ and $\Omega/ \pi^-$. This emphasizes the necessity of additional research to determine whether thermal hadron gas models are appropriate and stresses the significance of any necessary adjustments to improve their correctness.

\end{abstract}


\maketitle

\section{Introduction}
\label{sec:Intr}

Gaining an understanding of the hadron-parton phase diagram, which outlines different phases and transitions, is one of the main objectives of researching ultrarelativistic heavy-ion (HI) collisions \cite{banks1983deconfining}. The hadronic phase is well understood, in which stable baryons make up a large fraction of the Universe and ordinary events. Different phases do, however, appear at higher densities and/or temperatures. For example, chiral symmetry restoration and the transition to deconfinement occur around temperatures approaching 150 MeV \cite{tawfik2020deconfinement,ding2019chiral}. This has led to the hypothesis that quarks and gluons move almost freely within a colored phase known as the strongly coupled quark-gluon plasma (sQGP) \cite{shuryak2017strongly}. The presence of hadronic (baryonic) matter, which forms compact celestial structures such as neutron stars, is observed in conventional ways in low-temperature, high-density environments. Furthermore, recent reports have suggested the detection of gravitational waves from mergers of neutron stars \cite{LIGOScientific:2017vwq}. Extreme celestial bodies like quark stars may exist at much higher densities, according to some theories \cite{fraga2001small}. Two essential intensive parameters in equilibrium are temperature, $T$, and baryon chemical potential, $\mu_{\mathtt{b}}$. Quantum Chromodynamics (QCD) is a gauge field theory that describes the strong interactions between colored quarks and gluons as well as their color-neutral composite states. Strongly interacting matter at extremes of temperature and baryon chemical potential has been the focus of a large global theoretical and experimental endeavor. There are several stages of chiral and deconfinement transitions identified in Lattice Quantum Chromodynamics (LQCD), especially at low baryon concentrations.

Particle multiplicities and ratios can be used to systematically analyze the thermal properties of the final state in high-energy nuclear (HI) collisions due to the statistical features of particle generation. Large amounts of experimental data at energies ranging over four orders of magnitude of GeV have become available in the last several decades. There is evidence to suggest that a number of statistical thermal models \cite{Andronic:2005yp,braun1996thermal,Tawfik:2014eba,braun1999chemical,cleymans1993thermal,tiwari2013particle,rafelski2000sudden,becattini2001features,cleymans1998thermal} have been particularly effective in explaining the observed particle yields and their ratios in higher energy collisions. After analyzing a sizable dataset, it was determined that the generated particles mostly corroborate the theory that hadrons most likely arise from thermal sources at particular temperatures and chemical potentials. Except for some baryon-to-meson ratios, such as the proton-to-pion ratio at peak RHIC and LHC energies, which is called the proton anomaly \cite{tawfik2019equation}, it is clear that this thermal behavior is generic.

Two factors are effective in explaining the particle ratios under equilibrium conditions: the baryon chemical potential ($\mu_{\mathtt{b}}$) and the temperature of chemical freezeout ($T_{\mathtt{ch}}$). When inelastic reactions stop and the number of particles created is fixed in high-energy (HI) collisions, the chemical freezeout phase takes place. This phase is characterized by thermal models such as the ideal hadron resonance gas (IHRG) model, which outlines the basic properties of the hot and dense fireball produced in HI collisions. Thermal models, therefore, are an important link between the QCD phase diagram and HI experiments \cite{stephanov2005qcd,karsch2005thermodynamic}. Measurements, in particular, particle multiplicities, are correlated with number densities in thermal models, which in turn mainly depend on chemical freezeout parameters: $T_{\mathtt{ch}}$ and $\mu_{\mathtt{b}}$. The utilization of this method enables the application of thermodynamics, correlations, fluctuations, and other associated ideas to high-energy collisions \cite{Tawfik:2014eba}.

Typically, the hadronic sector in lattice QCD computations is measured against the ideal hadron resonance gas (IHRG) model \cite{Huovinen:2009yb,Borsanyi:2010bp}. At lower temperatures, there are some systematic differences seen, but overall, these calculations agree quite well with the IHRG model's predictions \cite{Tawfik:2014eba}. These differences may arise from the existence of new resonances that are not taken into account by the IHRG model, which is based on known resonances reported by the Particle Data Group \cite{Workman:2022ynf}, and potentially from the need to extend the model to include hadronic interactions.

According to recent studies in the literature \cite{samanta2019exploring,vovchenko2017multicomponent}, the traditional performance of the IHRG model appears to be inadequate to account for all of the information that is now available and the predictions from modern lattice QCD simulations. Numerous research has proposed integrating different types of interactions \cite{Tawfik:2004sw,Huovinen:2017ogf, Vovchenko:2017xad, Vovchenko:2017drx}. It is crucial to ascertain how to integrate interactions among the hadrons when comparing the thermodynamics computed within the thermal model framework to similar data obtained via lattice QCD methods.

Resonance formation is suggested by arguments originating from the S-matrix method \cite{dashen1969s,venugopalan1992thermal,lo2017s}, which state that the (ideal) IHRG model naturally includes attractive interactions between hadrons. Both repulsive and attractive forces among component hadrons are taken into consideration in more realistic hadronic interaction models. Repulsive interactions in the IHRG model were previously studied in the contexts of relativistic cluster and virial expansions \cite{venugopalan1992thermal}, through repulsive mean fields \cite{olive1981thermodynamics,olive1982quark}, and through excluded volume (EV) corrections \cite{hagedorn1980hot,gorenstein1981phase,hagedorn1983pressure,kapusta1982thermodynamics,rischke1991excluded,anchishkin1995generalization}. A substantial amount of research has been done on the effects of EV interactions among hadrons on heavy-ion collision observables \cite{braun1999chemical,cleymans2006comparison,begun2013hadron,fu2013higher,vovchenko2017examination,alba2018flavor,satarov2017new,vovchenko2017van} and on IHRG thermodynamics \cite{satarov2009equation,andronic2012interacting,bhattacharyya2014fluctuations,albright2014matching,vovchenko2015hadron,albright2015baryon,redlich2016thermodynamics,alba2017excluded}. Repulsive interactions have recently come back into focus due to lattice QCD findings on conserved charge fluctuations, suggesting that significant departures from the IHRG baseline of various fluctuation observables may be attributed to repulsive baryon-baryon interactions \cite{huovinen2018hadron,vovchenko2017van,vovchenko2017equations}.

A boundary of the chemical freezeout diagram is defined by the dependence of $\mu_{\mathtt{b}}$ and $T_{\mathtt{ch}}$ on the nucleon-nucleon center-of-mass energies $\sqrt{s_{\mathtt{NN}}}$, which closely resembles the QCD phase diagram \cite{braun1998dynamics}. The dependence of $T_{\mathtt{ch}}$ on $\mu_{\mathtt{b}}$ is similar to that of several thermodynamic quantities computed in lattice QCD \cite{stephanov2005qcd,karsch2005thermodynamic}. This is especially true at $\mu_b/T\leq 1$, that is, at $\sqrt{s_{\mathtt{NN}}}$ which is larger than top SPS energies. It follows that these bounds are still relevant \cite{tawfik2016possible}, especially at lower energies, that is, higher baryon chemical potentials, where QCD-like effective models are important, e.g., the IHRG model \cite{Tawfik:2014eba} and the Polyakov linear-sigma model (PLSM) \cite{tawfik2019equation,tawfik2018quark}. 

The Cosmic Ray Monte Carlo (CRMC) \cite{kalmykov1997quark,kalmykov1993nucleus,ostapchenko2006nonlinear,ostapchenko2006qgsjet,ostapchenko2007status,engel1992nucleus,fletcher1994s,ahn2009cosmic, werner2006parton,pierog2009epos} models are among the hybrid event generators that supply these frameworks. This study presents our calculations of particular particle ratios based on CRMC EPOS $1.99$ simulations. Along with our calculations based on our recently suggested quantum mechanically correlated statistical correction (HRG) to the IHRG model, we also compare these particle ratios with known experimental data and our equivalent calculations based on the ideal hadron gas (IHRG) model \cite{Hanafy:2020hgq}. In the next section, we will elaborate on our quantum-mechanically corrected HRG model and provide further justification beyond what we previously outlined in \cite{Hanafy:2020hgq}.

Apart from our main goal of further validating and evaluating our recently proposed statistical correction to the ideal IHRG model \cite{Hanafy:2020hgq}, the use of CRMC EPOS $1.99$ based data in this study allows us to make predictions about future facilities that will be used to explore the intermediate temperature region of the QCD phase diagram, such as the Nuclotron-based Ion Collider facility (NICA) at the Joint Institute for Nuclear Research (JINR), Dubna, Russia; the Facility for Antiproton and Ion Research (FAIR) at the Gesellschaft für Schwerionenforschung (GSI) in Darmstadt, Germany; the J-PARC heavy ion project (J-PARC-HI) at the Japan Proton Accelerator Research Complex (J-PARC) in Tokai, Japan; and the BES-II program at RHIC. In the meantime, low $\mu_{\mathtt{b}}$ or high $T_{\mathtt{ch}}$ are achieved by both the top RHIC and the LHC (on the left side of the typical QCD phase-diagram).

The remaining sections of this paper are organized as follows: The HRG model, a non-ideal (correlated) statistical correction to the IHRG model that draws inspiration from the Beth-Uhlenbeck (BU) quantum theory of non-ideal gases, is presented in Section \ref{sec:Mod} after a full formalism of the standard ideal (uncorrelated) IHRG model is reviewed. Lastly, we go over the simulation model used in this work—the EPOS event generator. We compare our calculations of specific particle ratios based on our previously suggested correction with the results obtained from the CRMC EPOS $1.99$ simulations and the IHRG model in Section \ref{sec:Res}. Next, a variety of experimental data on particle ratios are compared with all model-based computations. A summary and conclusion can be found in Section \ref{sec:Cncls}.

\section{Used Models}
\label{sec:Mod}

The idea of particle interaction probability is integrated in this study based on the work of Beth and Uhlenbeck on quantum-mechanical models for molecular interactions, as reported in reference \cite{uhlenbeck1936quantum, uhlenbeck1937quantum}. We began our endeavor in \cite{Hanafy:2020hgq} with the aim of suggesting a statistical modification to the uncorrelated (ideal) IHRG model.

A relationship was suggested by Beth and Uhlenbeck (BU) between the virial coefficients and the likelihood of discovering nearby particle pairs, triples, and so on. It was shown that under classical conditions, which are generally high temperatures and/or low particle densities, these probabilities (exact expressions to be discussed later in this section) can be expressed as Boltzmann factors provided that the de Broglie wavelength, a measure of quantum non-locality, stays small in relation to the particle's spatial extent, or the "diameter" \cite{uhlenbeck1936quantum, uhlenbeck1937quantum}. The term "diameter" is kept from the original terminology used by Beth and Uhlenbeck. The range of space in which a particle can interact with other particles in a fundamental (classical) manner is indicated by this particle diameter. After comparing the experimental results with the original BU model, it was determined that deviations from the classical excluded volume model due to quantum effects become significant at sufficiently low temperatures \cite{uhlenbeck1936quantum}. The particles considered in the original BU model were molecules such as helium. The thermal de Broglie wavelength approximates the particle diameter at this point.

The basic Beth and Uhlenbeck quantum mechanical model of particle interactions was extended in ref. \cite{uhlenbeck1936quantum} to include discrete quantum states for a general interaction potential (which need not be central) and the effect of Bose or Fermi statistics. The expression for the second virial expansion was then generalized using the cluster integral to describe particle interactions, provided that the particles did not form bound states \cite{dashen1969s,venugopalan1992thermal,kostyuk2001second}. This was done after the expression was developed in ref. \cite{uhlenbeck1936quantum} and then extended in ref. \cite{uhlenbeck1937quantum}.

Using the quantum mechanical BU technique, repulsive interactions between baryons in a hadron gas have been described more recently \cite{vovchenko2018beth}. \cite{vovchenko2018beth} determined the second virial coefficient, which is also referred to as the excluded volume parameter, within the BU technique. In addition to being temperature-dependent, it was found that this parameter deviated significantly from the outcome of the conventional excluded volume (EV) model. For nucleons at temperatures $T=100–200$ MeV, the widely used conventional EV model \cite{huang1963statistical,landau1980lifshitz,greiner2012thermodynamics} underestimated the EV parameter by large factors of 3-5, assuming a given value of the nucleon hard-core radius (expected to be $\simeq 0.3$ fm). Thus, it was concluded in \cite{vovchenko2018beth} that earlier studies that used hadron hard-core radii as inputs into the conventional EV model needed to be reassessed using appropriately rescaled quantum mechanical EV parameters.

We initially present the fundamental building blocks of the ideal IHRG model in the second part of this section. We then develop a statistical (HRG) model that is inspired by BU's quantum theory of non-ideal (correlated) gases \cite{uhlenbeck1937quantum}, as a correction to the ideal (uncorrelated) IHRG model. Therefore, a modified version of the partition function of the usual ideal IHRG is included in our computation formalism. Additionally, we offer a thorough explanation of the simulation model—the EPOS event generator—which we use to compare models.

\subsection{ Ideal HRG Model}

In the bootstrap framework \cite{fast1963statistical,fast1963nuovo,eden1966polkinghorne}, a partition function $Z(T, \mu, V)$ is associated with an equilibrium thermal model of an interaction-free gas. By computing derivatives suitably, one can deduce the thermodynamics of the system from this partition function. As stated in \cite{Tawfik:2014eba,Karsch:2003vd,Karsch:2003zq,Redlich:2004gp,Tawfik:2004sw,Tawfik:2005qh}, the partition function in a grand canonical ensemble.

\bea 
Z(T,V,\mu)=\mbox{Tr}\left[\exp\left(\frac{{\mu}N-H}{T}\right)\right], \label{eq:Z} 
\eea
whereby $N$ is the total number of members in the statistical ensemble and $H$ is the Hamiltonian incorporating all pertinent degrees of freedom. From the current particle data group (PDG) \cite{Workman:2022ynf}, all hadron resonances with masses up to $2.5~$GeV may be summed, yielding Eq. (\ref{eq:Z}).
\bea 
\ln  Z(T,V,\mu)=\sum_i{{\ln Z}_i(T,V,\mu)} = V \sum_i \frac{g_i}{2{\pi}^2}\int^{\infty}_0{\pm p^2 dp {\ln} {\left[1\pm {\lambda}_i \exp\left(\frac{-{\varepsilon}_i(p)}{T} \right) \right]}}, \label{eq:lnZ}
\eea
where the pressure can be denoted as $T\partial \ln  Z(T,V,\mu)/\partial V$, $\pm$ designates fermions and bosons, respectively. $\varepsilon_{i}=\left(p^{2}+m_{i}^{2}\right)^{1/2}$ is the dispersion relation, $g_i$ is the spin-isospin degeneracy factor, and $\lambda_i$ is the fugacity factor of the $i$-th particle \cite{Tawfik:2014eba},
\bea
\lambda_{i} (T,\mu)=\exp\left(\frac{B_{i} \mu_{\mathtt{b}}+S_{i} \mu_{\mathtt{S}}+Q_{i} \mu_{\mathtt{Q}}}{T} \right), \label{eq:lmbd}
\eea
where the quantum numbers of the $i$-th hadron are $B_{i} (\mu_{\mathtt{b}})$, $S_{i} (\mu_{\mathtt{S}})$, and $Q_{i} (\mu_{\mathtt{Q}})$ for baryons, strangeness, and electric charges, respectively, along with their corresponding chemical potentials. From a phenomenological perspective, the baryon chemical potential $\mu_{\mathtt{b}}$ can be related to the nucleon-nucleon center-of-mass energy $\sqrt{s_{\mathtt{NN}}}$ using the following parameterization: along the chemical freezeout boundary, where particle production is conjectured to cease \cite{tawfik2015thermal}.
\bea
\mu_{\mathtt{b}} &=& \frac{a}{1+b \sqrt{s_{\mathtt{NN}}}}, \label{eq:mue}
\eea
where
$a=1.245\pm0.049~$GeV and $b=0.244\pm0.028~$GeV$^{-1}$. Apart from pressure, the partition function can be used to directly determine the number and energy density, respectively, as well as the entropy density and other thermodynamics by calculating the appropriate derivatives.
\begin{eqnarray}
n_i(T,\mu) &=& 
\sum_{i}\frac{g_{i}}{2\pi^{2}}\int_{0}^{\infty}{p^{2} dp \frac{1}{\exp\left[\frac{\mu_{i} - \varepsilon_{i}(p)}{T}\right] \pm 1}}, \\ 
\rho_i(T,\mu) &=& 
\sum_{i}\frac{g_{i}}{2\pi^{2}}\int_{0}^{\infty}{p^{2} dp \frac{-\varepsilon_{i(p)}\pm \mu_i}{\exp\left[\frac{\mu_{i}-\varepsilon_i(p)}{T}\right] \pm 1}}. \label{eq:e}
\end{eqnarray}

It is crutial to appreciate that both $T$ and $\mu=B_{i} \mu_{\mathtt{b}}+S_{i} \mu_{\mathtt{S}}+\cdots$ are interrelated and linked to $\sqrt{s_{\mathtt{NN}}}$ \cite{Tawfik:2014eba}. Due to strangeness conservation, $\mu_{\mathtt{S}}$ is considered a dependent variable that needs to be determined, assuming an overall thermal equilibrium. In particular, the value allocated to $\mu_{\mathtt{S}}$ at a given $T$ and $\mu_{\mathtt{b}}$ guarantees that $\langle n_{\mathtt{S}}\rangle-\langle n_{\bar{\mathtt{S}}}\rangle=0$. In order to determine different thermodynamic quantities like particle number, energy, entropy, and so on, $\mu_{\mathtt{S}}$ is then combined with $T$ and $\mu_{\mathtt{b}}$. Functions of $T$, $\mu_{\mathtt{b}}$, and $\mu_{\mathtt{S}}$ may also be used to express the chemical potentials associated with other conserved quantum charges, such as the electric charge and the third component of isospin, and they all satisfy the corresponding conservation laws.

\subsection{The Modified HRG model}

As was briefly discussed in the section above, a two-particle interaction probability in the context of a quantum gas consisting of fermions and bosons with mass $m_{i}$ and correlation (interaction) distance $r$, at temperature $T$ and with zero $\mu_{\mathtt{b}}$, looks like this:

\bea
1\pm \exp\left(-4\pi^{2} m_{i} T r^{2}\right)
\eea 
The notion of correction the interactions of particles for quantum effects was first introduced by Beth and Uhlenbeck \cite{uhlenbeck1937quantum} in an attempt to characterize interactions inside a quantum gas of particles, assuming a generic potential and ignoring the development of bound states. Even for an ideal gas, which is a frequent approximation at high enough temperatures, the Boltzmann-like factor, $\exp\left(-4\pi^{2} {m}_{i} T r^{2}\right)$, is still relevant. Due to the shift in statistics, the apparent attraction or repulsion between bosons (fermions) is indicated by the $\pm$ sign. We introduce a modification to the probability term in the expression given in Eq. (\ref{eq:Z}) for the ideal hadron gas partition function, motivated by this correction. We provide a novel probability term that has the form
\bea
1\pm \lambda_i \exp\left(\frac{-{\varepsilon}_i(p)}{T}\right) \left[1\pm \exp\left(-4\pi^{2} {m}_{i} T r^{2}\right)\right].
\eea 

According to the BU quantum correlations framework \cite{uhlenbeck1937quantum}, this adjusted probability function distinctly integrates interactions inside the hadron resonance gas, where $r$ is the correlation (interaction) length between any two hadrons at equilibrium temperature $T$. We modify the non-correlated IHRG partition function $Z(T, \mu, V)$ to adopt the following form by applying our suggested corrected probability function to it:

\bea
\ln Z \textprime(T,V,\mu) = 
\sum_i V \frac{g_i}{2 {\pi}^2}\int^{\infty}_0{\pm p^2 dp {\ln} \left[\pm \lambda_{i} \exp\left(\frac{-{\varepsilon}_{i(p)}}{T} \right) \right] \left[1\pm \exp\left(4 \pi m_{i} T r^{2}\right)\right]}, \label{eq:lnZz} 
\eea

For the non-correlated IHRG scenario, this strategy basically consists of averaging all hadron resonances, using a similar method as described in the justification for Eq. (\ref{eq:lnZ}). Thus, by suitably differentiating $\ln Z \textprime$, the thermodynamic features of the correlated HRG model may be determined, thereby reproducing the earlier methods presented in the context of the non-correlated IHRG scenario.

\subsection{Cosmic Ray Monte Carlo (CRMC) model}
\label{sec:crmc} 

We report on the use of the hybrid event generator, Cosmic Ray Monte Carlo (CRMC EPOSlhc), to produce different particle ratios for different hadrons at energies between $\sqrt{s_{NN}}=3$ and $2760~$ GeV. The results obtained from the EPOSlhc event generator are next compared with existing experimental data and finally with our statistically corrected hadron resonance gas model (HRG) computations, which are inspired by quantum mechanics and are obtained from Eq. (\ref{eq:lnZz}).

The CRMC interface supports a variety of effective quantum chromodynamic (QCD) theories and experimental configurations, including NA61, ATLAS, CMS, LHCb, and the Pierre Auger Observatory for ultra-high-energy cosmic rays. Like the Gribov-Regge model, it includes a range of interactions based on the EPOSlhc/$1.99$ model and provides a backdrop description that takes into consideration the diffraction effects that arise. Additionally, the interface can be used to access outputs from multiple heavy-ion collision event generators and is compatible with a number of spectrum models, such as the EPOS $1.99$/lhc \cite{werner2006parton, pierog2009epos}, qgsjetII \cite{ostapchenko2006qgsjet, ostapchenko2006nonlinear, ostapchenko2007status}, qgsjet$01$ \cite{kalmykov1997quark, kalmykov1993nucleus}, and sibyll \cite{engel1992nucleus, fletcher1994s, ahn2009cosmic}. At ultra-relativistic high energies, QGSJETII, v$03$, v$04$, and EPOS lhc/$1.99$ are effective models, while SIBYLL$2.3$ and QGSJET$01$ are used at low energies.

We used the EPOSlhc event generator in this work, which includes a range of parameters for critical high-energy heavy-ion collision observables and associated phenomenological assumptions. Theoretical and experimental factors can be used to modify these parameters. Based on data gathered from many studies and other event generators, EPOSlhc has been praised for offering a realistic representation of heavy-ion collisions \cite{werner2006parton, pierog2009epos}.

EPOSlhc was initially developed for cosmic ray air showers, but it can also be used for pp- and AA-collisions at energies from SPS to LHC. It utilizes reduced-order treatments for heavy-ion collisions close to the end of their development, allowing for the least amount of bias in hadron interactions in nuclear collisions \cite{pierog2015epos}. EPOSlhc functions as a parton model, combining many parton-parton interactions that result in various parton ladders. Particle production, multiple parton scattering, cross-section measurements, splitting-based shadowing and screening, and different combined effects of hot and dense matter are all provided with robust estimates.

We used the EPOSlhc event generator in this paper to simulate at least $100,000$ events at each energy level, ranging from $3$ to $2760~$ GeV. Particle ratios were calculated for a number of particle yields: equal mass ($\bar{p} / p$, $K^-/ K^+$, $\pi^-/ \pi^+$, $\bar{\Lambda}/ \Lambda$, $\bar{\Sigma}/ \Sigma$, $\bar{\Omega}/ \Omega$) and unequal mass ($p/ \pi^+$, $k^+/ \pi^+$, $k^-/ \pi^-$, $\Lambda/ \pi^-$, $\bar{p}/ \pi^-$, $\Omega/\pi^-$) over a broad rapidity window ($-6 < y < 6$). We sought to investigate the particle ratios for various hadron species in order to improve our comprehension of the validity of the hybrid EPOSlhc event generator. It is anticipated that this investigation will yield insightful information about how to improve these event generators, assisting efforts to gather new data pertinent to upcoming experimental facilities like NICA and FAIR.

\section {Results and Discussion}
\label{sec:Res}
  
Below are the comparison figures between the particle ratio data from our statistically corrected hadron resonance gas model (HRG) inspired by quantum mechanics, using Eq. (\ref{eq:lnZz}) (black solid curves) and our calculations of the same particle ratios using the ideal hadron resonance gas model (IHRG) based on Eq. (\ref{eq:lnZ}) (red dashed curves). Furthermore, we give our comparable calculations using CRMC EPOS $1.99$ simulations (blue stars), and compare them with related experimental results (different symbols). These experimental results are based in part on observations collected at the Superproton Synchrotron (SPS) \cite{braun1996thermal}, the Alternating Gradient Synchrotron (AGS) \cite{braun1999chemical}, NA49 \cite{anticic2004lambda,afanasiev2002xi, alt2005omega}, NA44 \cite{anticic2004lambda,afanasiev2002xi, alt2005omega, afanasiev2002energy, blume2005review}, NA57 \cite{antinori2004energy}, and PHENIX \cite{Adler:2006xd}. Future facilities like NICA and FAIR \cite{Taranenko:2020vqn,Galatyuk:2019lcf}, operating within the temperature range $T\in [130, 200~$MeV$]$, are expected to have access to a large chunk of this beam energy range. These temperatures are notably typical for the QCD equation-of-state and for phenomenological applications in heavy-ion collisions. We adjusted the quantum-mechanical correlation length (parameter) to $r$ = 0.85 fm, which is the best value, in all of our computations using our new HRG model.  

\begin{figure}[htb]
\includegraphics[width=7cm]{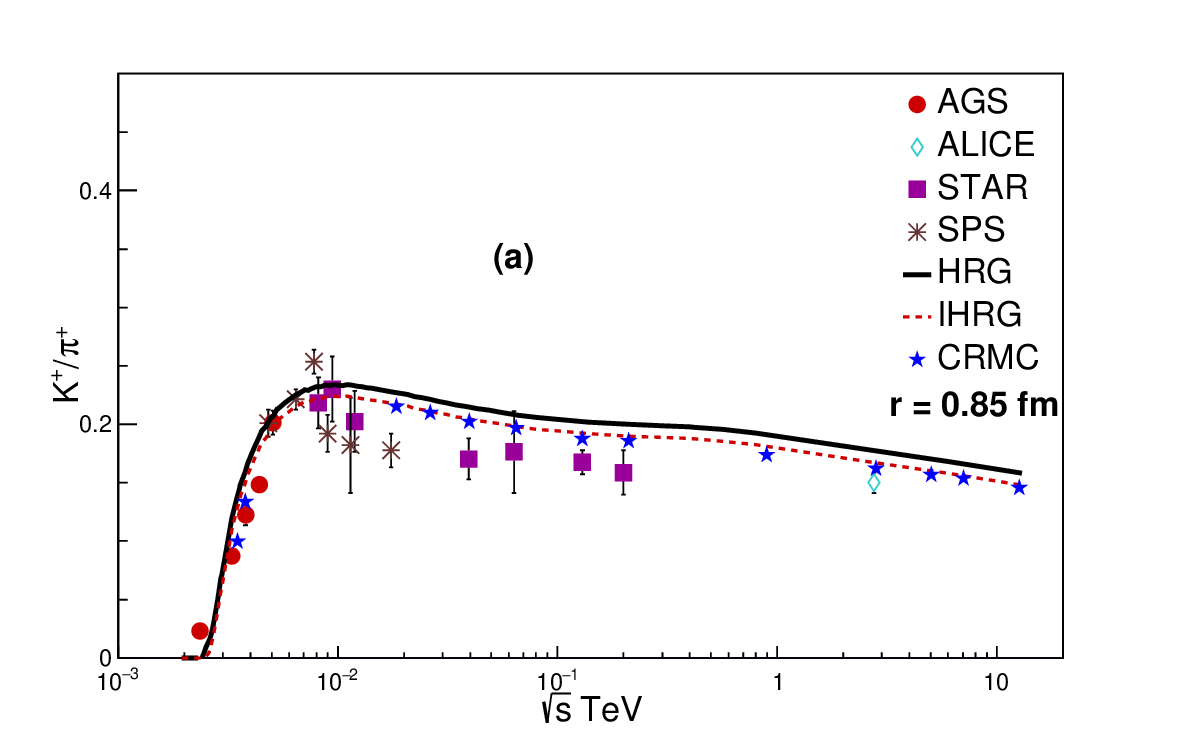}
\includegraphics[width=7cm]{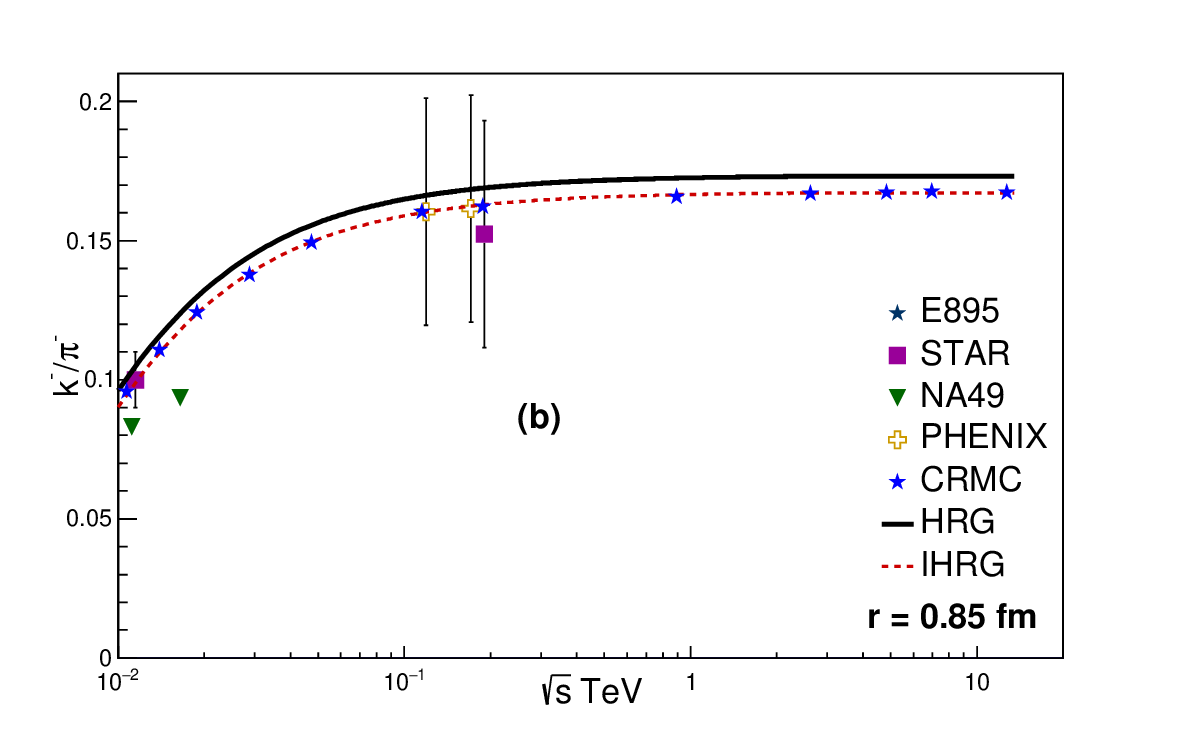} 
\includegraphics[width=7cm]{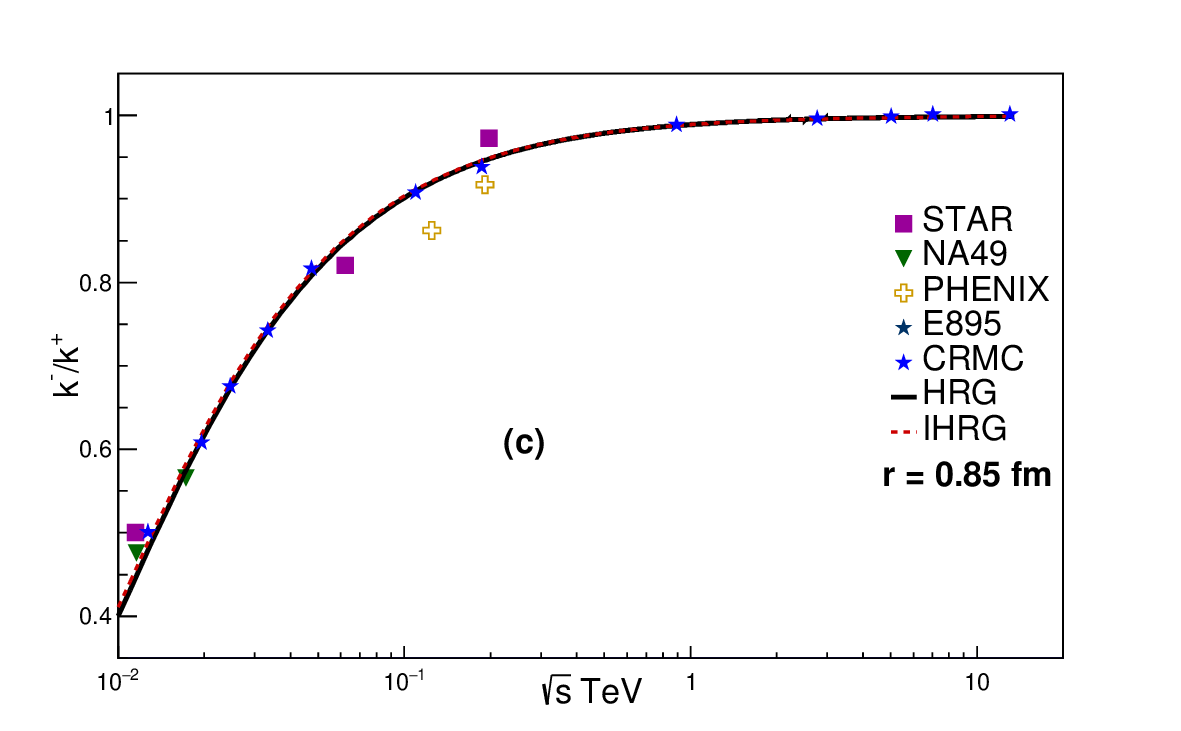} 
\includegraphics[width=7cm]{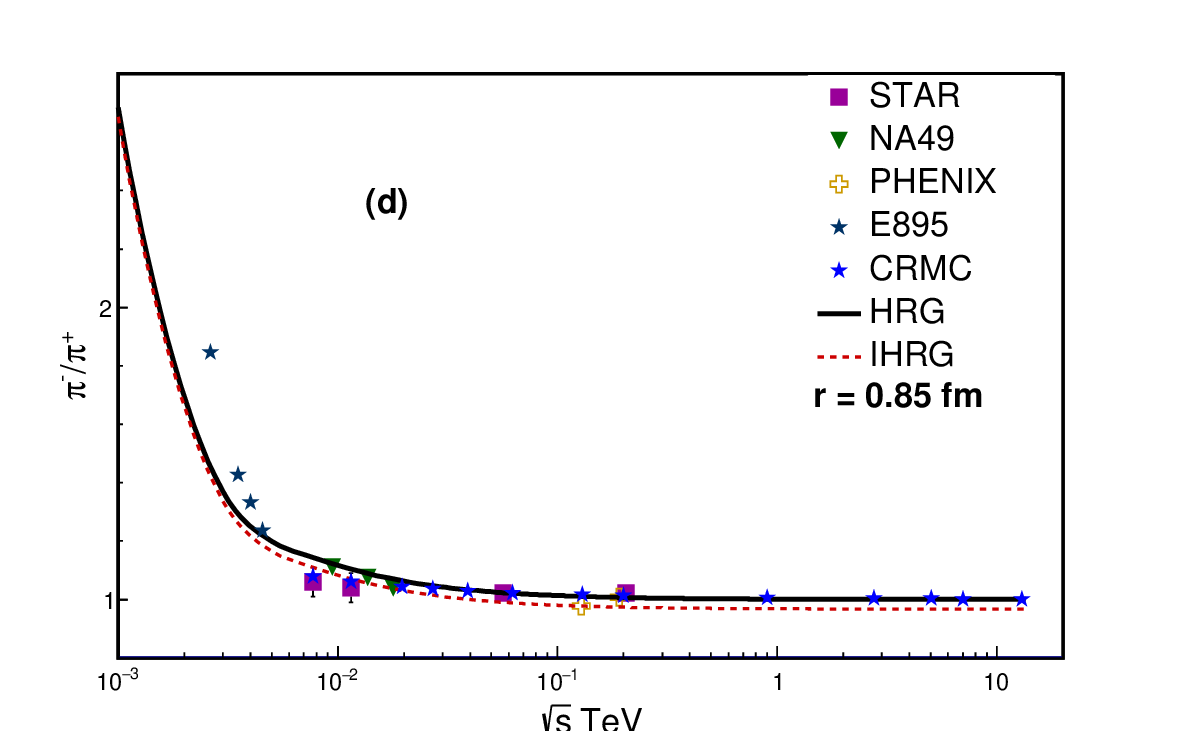} 

\caption {Variation of the strange meson to non-strange meson ratios $k^+/\pi^+$ (a), $k^-/\pi^-$ (b), antikaon-to-kaon ratio $k^-/k^+$ (c), and antipion-to-pion ratio $\pi^-/\pi^+$ (d)  with respect to center-of-mass energy $\sqrt{s_{\mathtt{NN}}}$.  Black solid curve, red dashed curve, and blue stars show the predictions of our new (HRG) model (Eq. \ref{eq:lnZz}), the ideal IHRG model (Eq. \ref{eq:lnZ}), and CRMC EPOS $1.99$ simulations \cite{werner2006parton, pierog2009epos}, respectively. The corresponding experimental data are represented by different symbols 
\cite{Bzdak:2019pkr,braun1999chemical,braun1996thermal,anticic2004lambda,afanasiev2002xi,alt2005omega,anticic2004lambda,afanasiev2002xi, alt2005omega,afanasiev2002energy,blume2005review,antinori2004energy,Adler:2006xd,Adam:2015kca}. }

\label{fig:one (a)–(d)}
\end{figure}

In Figs. \ref{fig:one (a)–(d)}, we show the center-of-mass energy dependence of the ratio of strange meson to non-strange meson such as $k^+/\pi^+$ (a), $K^-/ \pi^-$ (b), and the ratio of antimeson-to-meson such as $k^-/ k^+$ (c), and $\pi^-/ \pi^+$ calculated using our new HRG model defined in Eq.\ref{eq:lnZz} (Black solid curve), the ideal IHRG model defined in Eq.\ref{eq:lnZ} (red dashed curve), and CRMC EPOS $1.99$ simulations (blue stars) \cite{werner2006parton, pierog2009epos}. The corresponding experimental data of the respective particle ratios used in the comparison are represented by different symbols  \cite{Bzdak:2019pkr,braun1999chemical,braun1996thermal,anticic2004lambda,afanasiev2002xi,alt2005omega,anticic2004lambda,afanasiev2002xi, alt2005omega,afanasiev2002energy,blume2005review,antinori2004energy,Adler:2006xd,Adam:2015kca}. 

Similar to other variants of hadron thermal models, our novel quantum-mechanically inspired statistically corrected HRG model (illustrated by the red dashed curve) exhibits a slight tendency to overestimate the experimental $k^+/ \pi^+$ data, as depicted by various symbols in (Fig. 1a), from approximately $\sqrt{s_{\mathtt{NN}}}$ = 3 GeV onwards. Nonetheless, our updated HRG model (red dashed curve) shows a closer alignment with the experimental data compared to the ideal HRG model (black solid curve). Notably, both our new IHRG model (red dashed curve) and the HRG model (black solid curve) effectively capture the well-known peak around $\sqrt{s_{\mathtt{NN}}}$ = 10 GeV in the $k^+/ \pi^+$ ratio profile (Fig. 1a). In contrast to the experimental data, the calculations from our new IHRG model and the ideal HRG model for the $k^+/ \pi^+$ ratio appear to exhibit broader peaks. Commencing from the peak, our IHRG model demonstrates the closest fit to the available experimental data, particularly in the high $\sqrt{s_{\mathtt{NN}}}$ range, where our model closely matches the 2.76 TeV ALICE measurement \cite{Adam:2015kca} of the $k^+/ \pi^+$ ratio. Conversely, our CRMC EPOS $1.99$ simulations (depicted by blue stars) for the $k^+/ \pi^+$ data display significant suppression and notably fail to reproduce the prominent peak in the $k^+/\pi^+$ experimental data. However, they gradually approach the performance of our IHRG model and the ideal HRG model, closely fitting the experimental data from approximately $\sqrt{s_{\mathtt{NN}}}$ = 100 GeV up to the very high $\sqrt{s_{\mathtt{NN}}}$ limit (approximately 10 TeV).

For the $k^-/ \pi^-$ experimental data (depicted in Fig. 1b), our new IHRG model (red dashed curve) generally exhibits a very close agreement with the experimental data, slightly outperforming the ideal HRG model (black solid curve). Conversely, our CRMC EPOS $1.99$ simulations for the $k^-/ \pi^-$ experimental data capture the consistent upward trend observed in the $k^-/ \pi^-$ ratio (Fig. 1b), albeit with significant suppression across the energy range investigated in this study. However, this suppression in the CRMC EPOS $1.99$ simulations gradually diminishes in the TeV scale.

In Fig. 1c, the variation of the ratio $k^-/k^+$ with $\sqrt{s_{\mathtt{NN}}}$ is illustrated. Our new IHRG model's prediction for the $k^-/k^+$ profile (red dashed curve) closely mirrors that of the ideal HRG model (black solid curve). For $k^-/k^+$, all the featured experimental data exhibit excellent agreement with both our IHRG model and the ideal HRG model. Only the PHENIX measurement at lower energies near $\sqrt{s_{\mathtt{NN}}}$ = 100 GeV \cite{Adler:2006xd} appears to be slightly overestimated by all thermal models considered in this study. Our CRMC EPOS $1.99$ simulations for the $k^-/k^+$ ratio effectively capture the monotonically increasing trend observed in the experimental data up to the very high energy limit (approximately 10 TeV). In the low energy range up to about $\sqrt{s_{\mathtt{NN}}}$ = 30 GeV, CRMC EPOS $1.99$ simulations align well with both the thermal models and the available experimental data for the $k^-/k^+$ ratio. This alignment persists in the high energy range starting from about $\sqrt{s_{\mathtt{NN}}}$ = 3 TeV. However, between these two extremes in the intermediate energy regime, our CRMC simulations do not align well with either the experimental or the thermal model data.

Fig. 1d illustrates the variation of the ratio $\pi^-/ \pi^+$ with $\sqrt{s_{\mathtt{NN}}}$. For $\pi^-/ \pi^+$, all the featured experimental data exhibit strong agreement with both our IHRG model (red dashed curve) and the ideal HRG model (black solid curve), except for the lower energy E895 measurement data near $\sqrt{s_{\mathtt{NN}}} \simeq 2$ GeV \cite{Lisa:2005vw}, which are underestimated by all thermal models considered in this study. Our CRMC simulations align well with the considered $\pi^-/ \pi^+$ experimental data as well as with the featured thermal models starting from approximately 10 GeV up to the very high energy range (approximately 10 TeV).

\begin{figure}[htb]
\includegraphics[width=7cm]{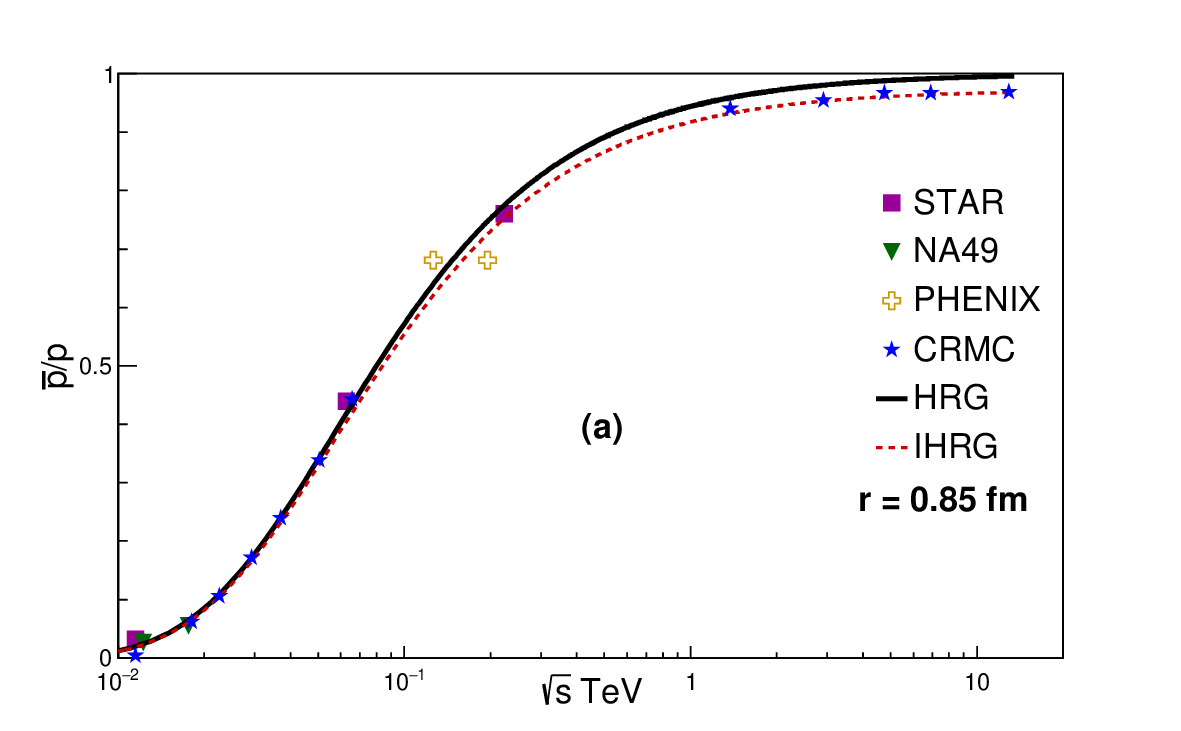}
\includegraphics[width=7cm]{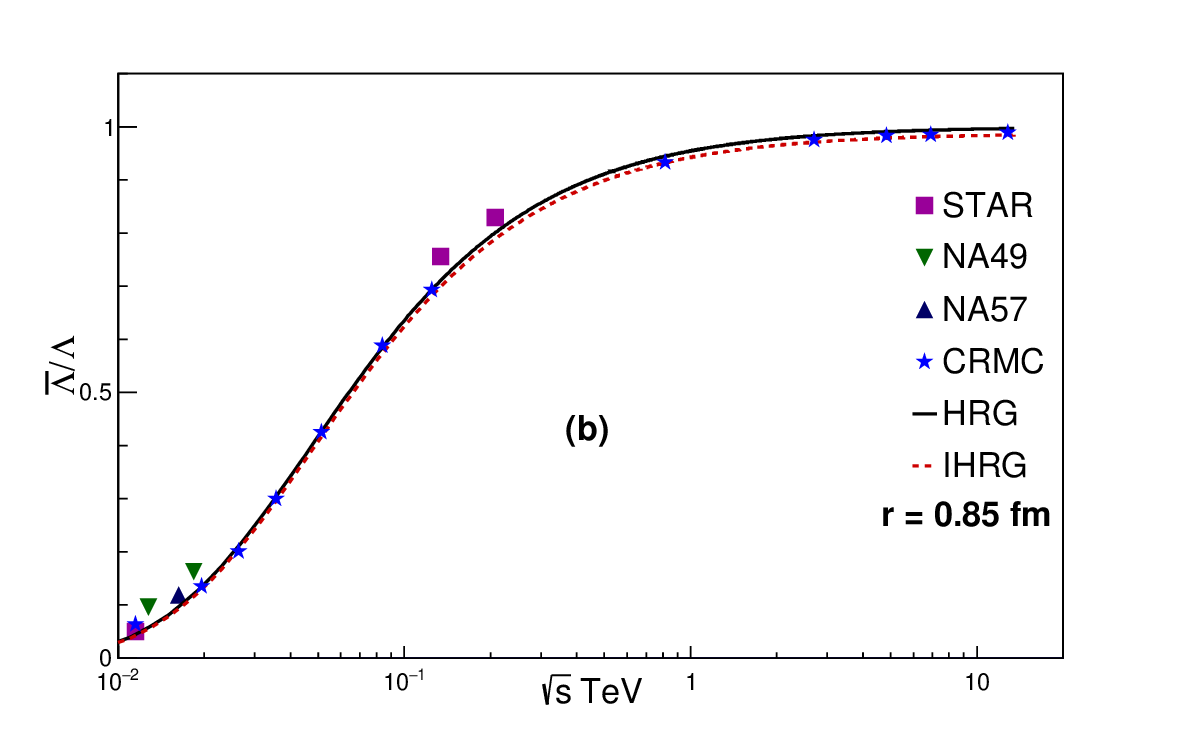} 
\includegraphics[width=7cm]{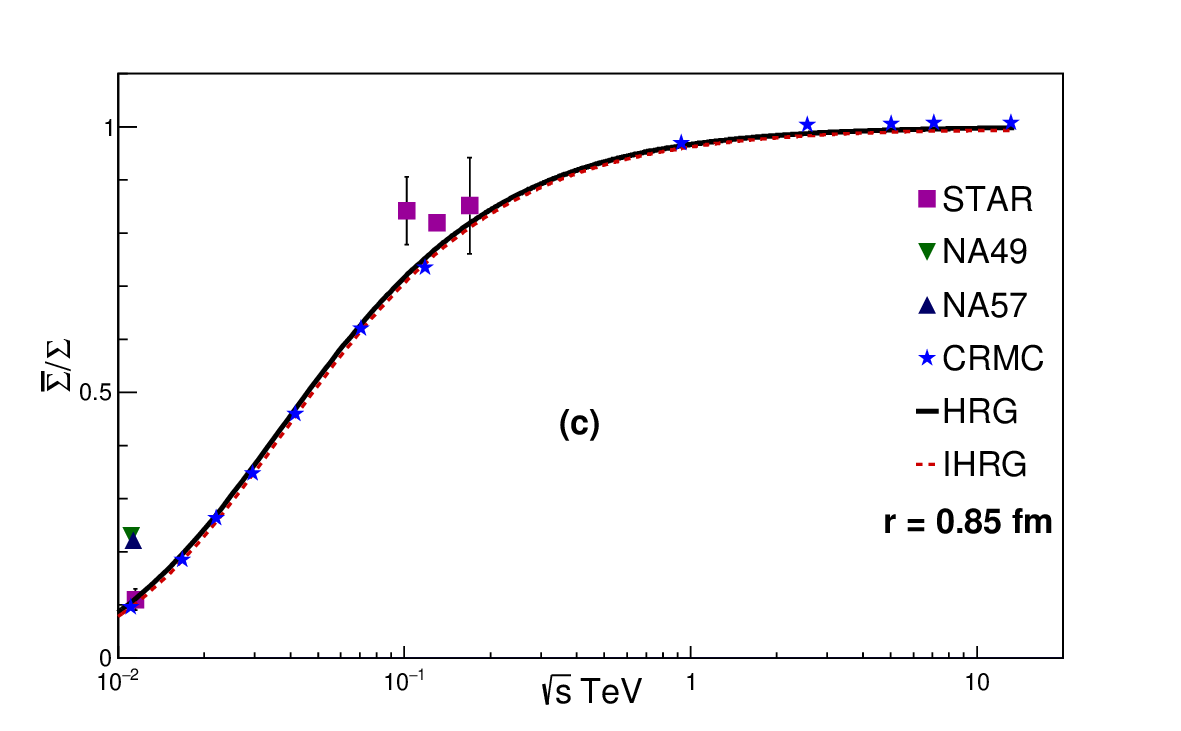} 
\includegraphics[width=7cm]{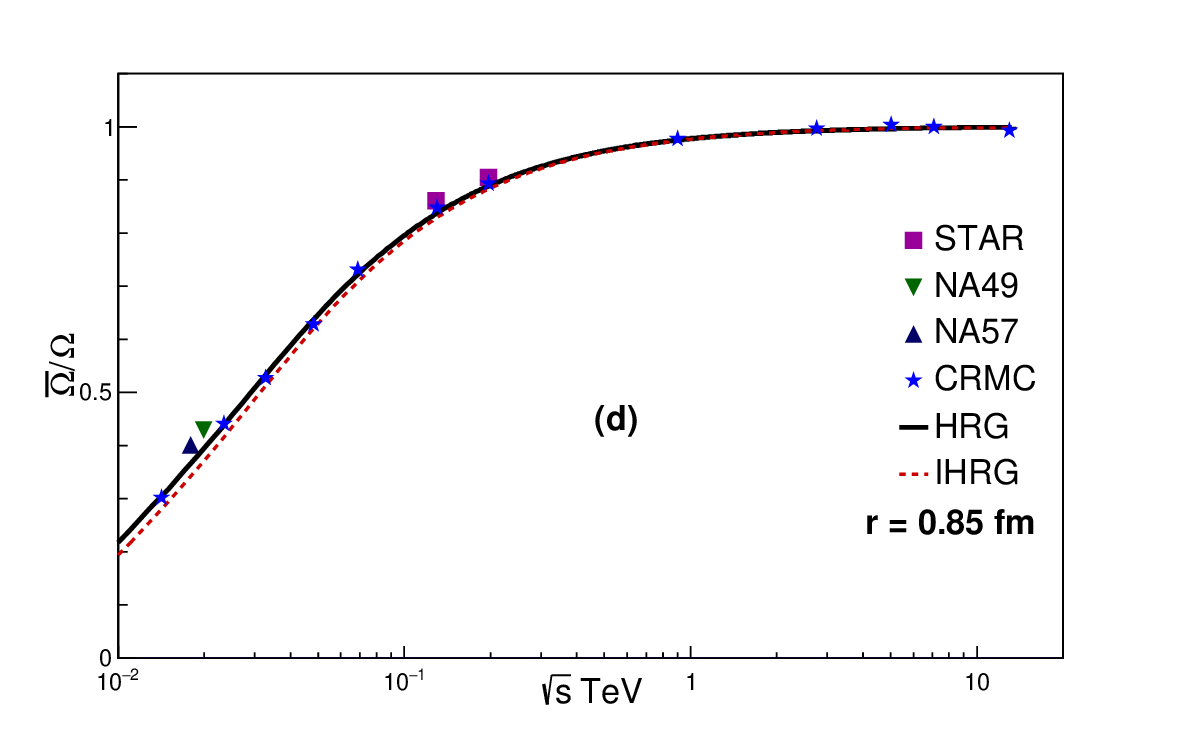} 

\caption{Variations of the antibaryon-to-baryon ratios such as $\bar{p}/p$ (a), $\bar{\Lambda}/ \Lambda$ (b),
 $\bar{\Sigma}/\Sigma$ (c), and  $\bar{\Omega}/\Omega$ (d) with respect to center-of-mass energy $\sqrt{s_{\mathtt{NN}}}$. red dashed curve, Black solid curve, and blue stars (curve) show the predictions of our new IHRG model (Eq. \ref{eq:lnZz}), ideal HRG model (Eq. \ref{eq:lnZ}), and CRMC EPOS $1.99$  \cite{werner2006parton, pierog2009epos}, respectively. The corresponding experimental points are represented by different symbols  \cite{Bzdak:2019pkr,braun1999chemical,braun1996thermal,anticic2004lambda,afanasiev2002xi,alt2005omega,anticic2004lambda,afanasiev2002xi, alt2005omega,afanasiev2002energy,blume2005review,antinori2004energy,Adler:2006xd,Adam:2015kca,back2004production}}.

\label{fig:two (a)–(d)}
\end{figure}   

In Figure \ref{fig:two (a)–(d)}, we show the center-of-mass energy dependence of the antibaryon-to-baryon ratios such as $\bar{p}/p$ (a), $\bar{\Lambda}/\Lambda$ (b), $\bar{\Sigma}/\Sigma$ (c), and $\bar{\Omega}/\Omega$ (d) calculated using our new HRG model defined in Eq. \ref{eq:lnZz} (Black solid curve), the ideal IHRG model defined in Eq. \ref{eq:lnZ} (red dashed curve), and CRMC EPOS $1.99$ simulations (blue stars curve) \cite{werner2006parton, pierog2009epos}. The corresponding experimental data of the respective particle ratios used in the present study are represented by different symbols  \cite{Bzdak:2019pkr,braun1999chemical,braun1996thermal,anticic2004lambda,afanasiev2002xi,alt2005omega,anticic2004lambda,afanasiev2002xi, alt2005omega,afanasiev2002energy,blume2005review,antinori2004energy,Adler:2006xd,Adam:2015kca,back2004production}. 

For the $\bar{p}/p$ ratio (Fig. 2a), it's evident that both our new HRG model and the ideal IHRG model exhibit very good agreement with experimental data up to the 200 GeV RHIC data, which represents the highest energy available experimentally. Conversely, our CRMC simulations notably underestimate the experimental data, especially for energies ranging from approximately 20 GeV to 200 GeV. Up to around $\sqrt{s_{\mathtt{NN}}} \simeq 200$ GeV, the $\bar{p}/p$ ratio calculated by both our new HRG model and the ideal IHRG model is nearly indistinguishable.

On the other hand, the profiles of the ratios $\bar{\Lambda}/\Lambda$ (Fig. 2b) and $\bar{\Sigma}/\Sigma$ (Fig. 2c) exhibit almost all the significant features outlined in discussing the $\bar{p}/p$ ratio (Fig. 2a), except for the notable observation that the experimental NA49 data at low energies (around $\simeq$ 10 GeV) generally appear to be underestimated by our thermal model calculations compared to the $\bar{p}/p$ ratio profile (Fig. 2a).

Our CRMC EPOS $1.99$ simulations notably underestimate the experimental data across the entire energy range in the $\bar{\Lambda}/\Lambda$ profile (Fig. 2b), and the same trend is observed for the $\bar{\Sigma}/\Sigma$ profile (Fig. 2c), except for the very high energy limit ($\sqrt{s_{\mathtt{NN}}} \gtrapprox$ 1 TeV), where the CRMC EPOS $1.99$ simulations clearly overestimate the thermal model calculations for the $\bar{\Sigma}/\Sigma$ ratio. Unfortunately, to our knowledge, no relevant experimental data are available in this energy regime.

In Figure 2d, the profile of the ratio $\bar{\Omega}/\Omega$ based on the thermal model calculations not only underestimates the experimental data at low energies (around $\simeq$ 10 GeV) but also at medium energies near ($\simeq$ 200 GeV). Here, our CRMC EPOS $1.99$ simulations appear to overestimate the thermal model calculations in the low energy regime, while underestimating the experimental data near the 200 GeV RHIC data. Moreover, our HRG model calculations (black solid curve) for this ratio are nearly identical to the ideal IHRG model calculations (red dashed curve).

\begin{figure}[htb]
\includegraphics[width=7cm]{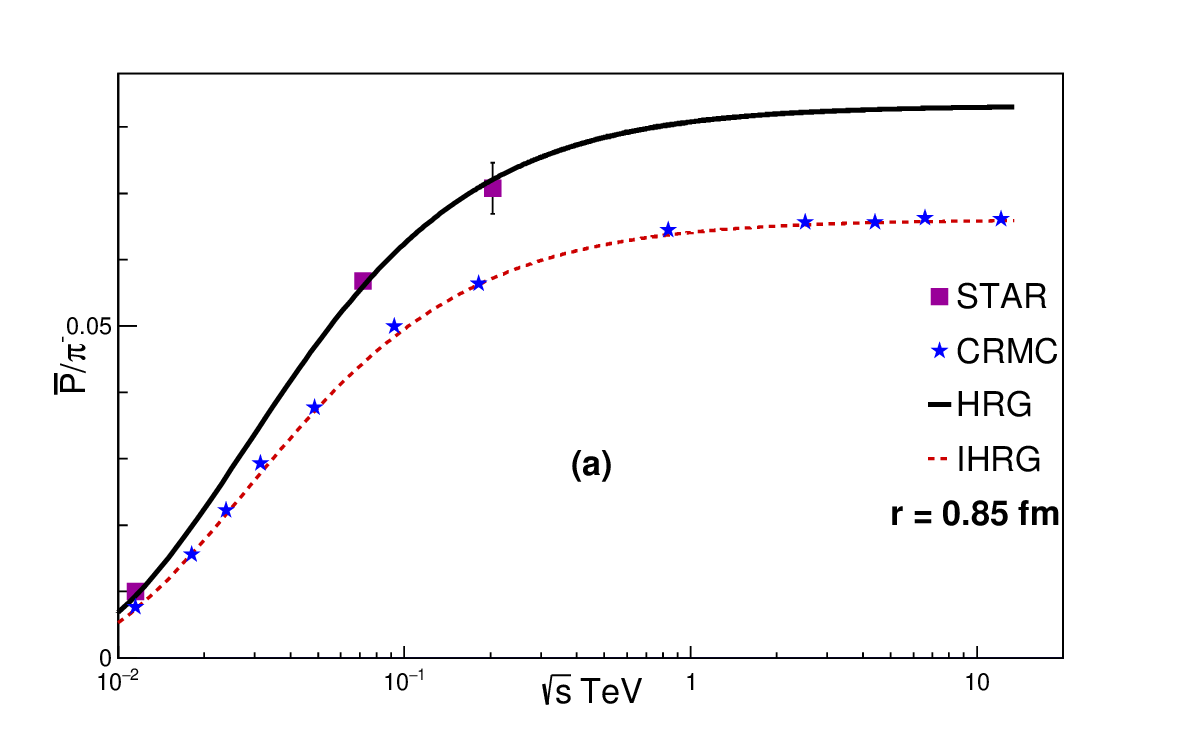}
\includegraphics[width=7cm]{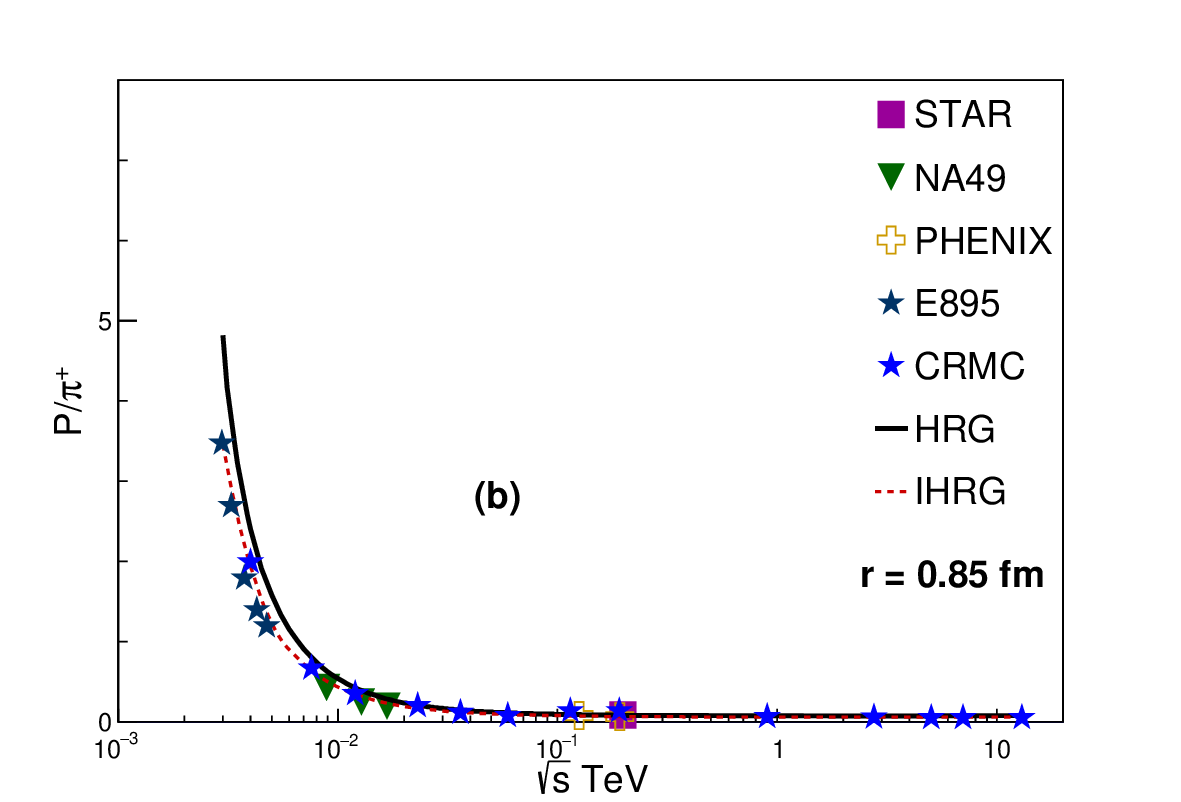} 
\includegraphics[width=7cm]{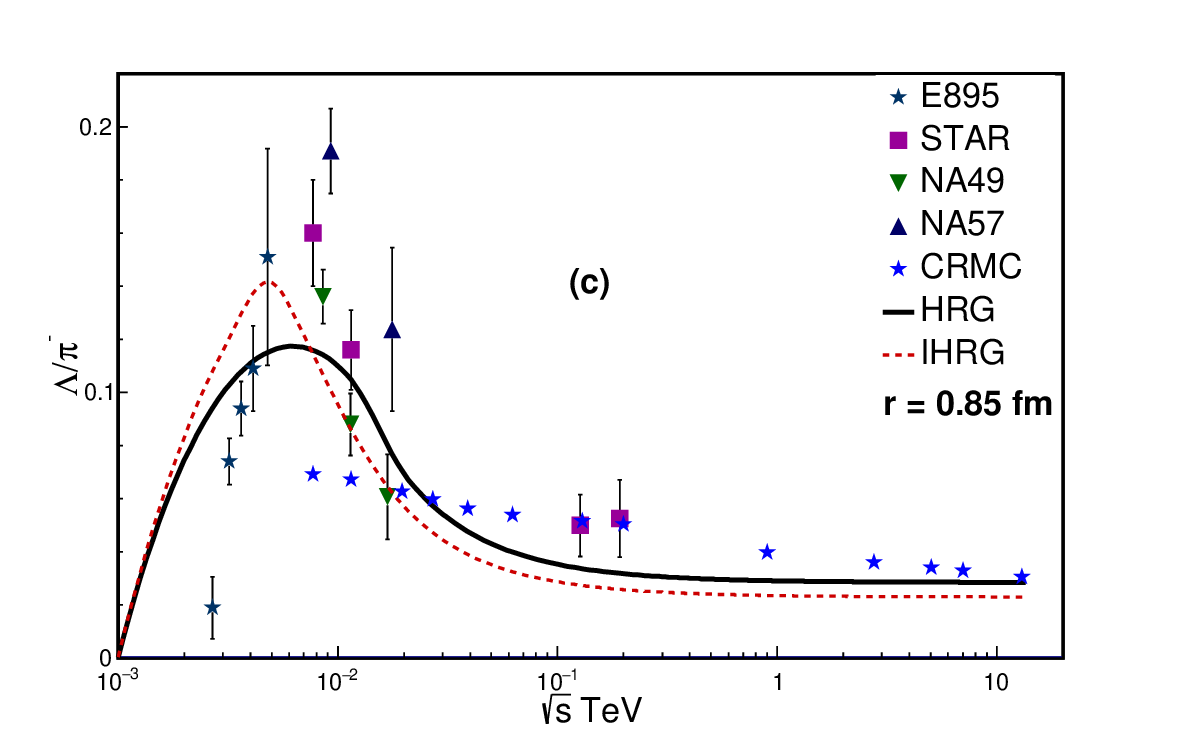} 
\includegraphics[width=7cm]{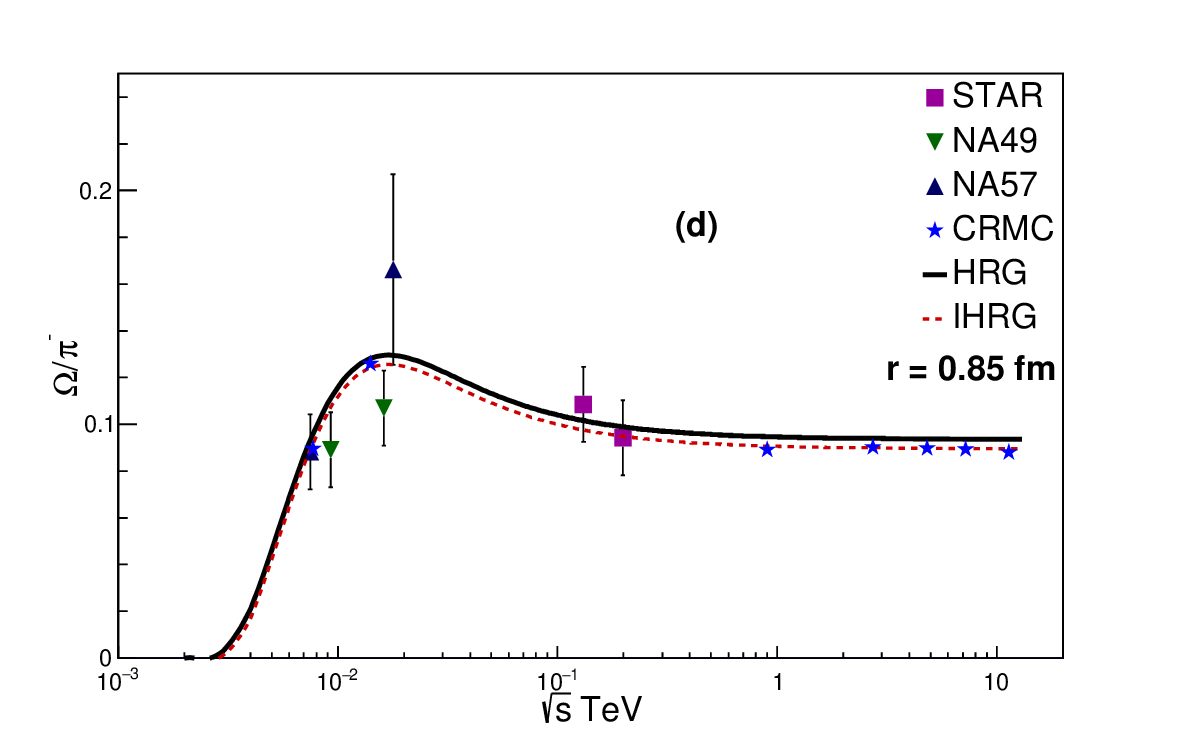} 

\caption{Variations of the antibaryon-to-antimeson ratios such as $\bar{p}/\pi^-$ (a), and baryon-to-meson ratios such as $p/\pi^+$ (b), $\Lambda/\pi^-$ (c), and $\Omega/\pi^-$ (d) with respect to center-of-mass energy $\sqrt{s_{\mathtt{NN}}}$. Black solid curve, red dashed curve, and blue stars show the predictions of our new HRG model (Eq. \ref{eq:lnZz}), ideal IHRG model (Eq. \ref{eq:lnZ}), and CRMC EPOS $1.99$ simulations  \cite{werner2006parton, pierog2009epos}, respectively. The corresponding experimental data are represented by different symbols  \cite{Bzdak:2019pkr,braun1999chemical,braun1996thermal,anticic2004lambda,afanasiev2002xi,alt2005omega,anticic2004lambda,afanasiev2002xi, alt2005omega,afanasiev2002energy,blume2005review,antinori2004energy,Adler:2006xd,Adam:2015kca,back2004production}}.

\label{fig:three (a)–(d)}
\end{figure}   

In Figure \ref{fig:three (a)–(d)}, we show the center-of-mass energy dependence of the antibaryon-to-antimeson ratios such as $\bar{p}/\pi^-$ (a), baryon-to-meson ratios such as $p/\pi^+$ (b), and baryon-to-antimeson ratios such as $\Lambda/\pi^-$ (c) and $\Omega/\pi^-$ (d) with respect to center-of-mass energy $\sqrt{s_{\mathtt{NN}}}$  calculated using our new HRG model defined in Eq. \ref{eq:lnZz} (black solid curve)), the ideal IHRG model defined in Eq. \ref{eq:lnZ}  (red dashed curve), and CRMC EPOS $1.99$ simulations (blue stars) \cite{werner2006parton, pierog2009epos}. The corresponding experimental data of the respective particle ratios used in the present study are represented by different symbols \cite{Bzdak:2019pkr,braun1999chemical,braun1996thermal,anticic2004lambda,afanasiev2002xi,alt2005omega,anticic2004lambda,afanasiev2002xi, alt2005omega,afanasiev2002energy,blume2005review,antinori2004energy,Adler:2006xd,Adam:2015kca,back2004production}. 

For the $\bar{p}/\pi^-$ ratio (Fig. 3a), it is evident that our new HRG model aligns very well with the available STAR experimental data points up to 200 GeV, whereas the ideal IHRG model consistently falls short of reproducing the experimental data, except at very low energies where all thermal model predictions converge with the lowest energy STAR data. Similar to the IHRG data, the CRMC EPOS $1.99$ simulations notably underestimate the experimental data, particularly at high RHIC energies near 200 GeV and extending to the high-energy limit ($\lesssim$ 10 TeV).

The $p/\pi^+$ ratio (Fig. 3b) shares nearly all the significant features outlined in discussing the $\bar{p}/\pi^-$ ratio (Fig. 3a), except that in this instance, the calculations based on our new HRG model only slightly outperform the corresponding ideal HRG model calculations up to high RHIC energies near 200 GeV.

In Fig. 3c and Fig. 3d, our new IHRG model calculations and the ideal HRG model calculations faithfully replicate the $\Lambda/\pi^-$ (Fig. 3c) and $\Omega / \pi^-$ (Fig. 3d) main features of the experimental profiles, particularly the peaks near $\sqrt{s_{\mathtt{NN}}}$ $\backsimeq$ 2 GeV in the $\Lambda / \pi^-$ profile and near $\sqrt{s_{\mathtt{NN}}}$ $\backsimeq$ 20 GeV in the $\Omega/\pi^-$ profile. Our IHRG model offers a highly accurate prediction of the position and shape of the $\Lambda / \pi^-$ ratio profile peak (Fig. 3c). The sharpness and height of our model's prediction for the peak of the $\Lambda/ \pi^-$ ratio is notably superior to that of the ideal HRG model. Concerning the $\Omega/\pi^-$ profile (Fig. 3d), our new HRG model and the ideal IHRG model perform nearly equally well in predicting the shape and position of the peak at about $\sqrt{s_{\mathtt{NN}}}$ $\backsimeq$ 20 GeV. However, the calculations based on our new HRG model and the ideal IHRG model appear to underestimate the height of the $\Lambda/\pi^-$ and $\Omega/\pi^-$ peaks. Finally, our CRMC EPOS $1.99$ simulations do not seem to reproduce the $\Lambda/\pi^-$ peak at all, while they produce the $\Omega/\pi^-$ peak but with a significantly exaggerated height.

\section{Summary and Conclusion}
\label{sec:Cncls}

We compared the particle ratios across different center-of-mass energies $\sqrt{s_{\mathtt{NN}}}$ using our novel quantum-mechanically inspired and statistically corrected hadron resonance gas (HRG) model, as defined by Eq. (\ref{eq:lnZz}), alongside calculations from the ideal hadron resonance gas model (IHRG) based on Eq. (\ref{eq:lnZ}). We also included calculations derived from CRMC EPOS $1.99$ simulations, and experimental data from STAR BES-I \cite{Bzdak:2019pkr}, as well as data partially gathered from AGS \cite{braun1999chemical}, SPS \cite{braun1996thermal}, NA49 \cite{anticic2004lambda,afanasiev2002xi, alt2005omega}, NA44 \cite{anticic2004lambda,afanasiev2002xi, alt2005omega, afanasiev2002energy, blume2005review}, NA57 \cite{antinori2004energy}, and PHENIX \cite{Adler:2006xd}. Much of this energy range is anticipated to be explored by future facilities like NICA and FAIR \cite{Taranenko:2020vqn,Galatyuk:2019lcf}, typically within the temperature range $T \in$ [$130, 200~$MeV], which is relevant for various applications in heavy-ion collisions and QCD equation-of-state studies. It's worth noting that we previously compared our new HRG model with lattice data in a separate study \cite{Hanafy:2020hgq}.

Overall, our new HRG model demonstrates a strong agreement with experimental data, particularly for particle ratios such as $k^-/k^+$, $\bar{\Lambda}/\Lambda$, $\bar{\Sigma}/\Sigma$, $\bar{\Omega}/\Omega$, $\Omega/\pi^-$, and $\pi^-/\pi^+$. While some particle ratios show less alignment with our model, it is noteworthy that our HRG model performs exceptionally well for the $\bar{p}/\pi^-$ and $p/\pi^+$ pairs. This suggests potential for addressing the proton anomaly issue at top RHIC and LHC energies \cite{tawfik2019equation}, although further investigation is required for confirmation.

While the ideal IHRG model also demonstrates generally good agreement with experimental data, the incorporation of repulsive interactions among hadrons represents a significant improvement in our new HRG model.

In conclusion, the results presented in our study support the efficacy of our new quantum-mechanically correlated and statistically corrected HRG model in elucidating particle ratios post-chemical freeze-out.

\section{References}

\bibliographystyle{aip}

\bibliography{Quantum_Correction_with_Ratios}

\end{document}